%% file: chiralOpti.tex
\documentclass[a4paper]{jpconf}
\usepackage{graphicx}
\input{preamble}

\begin{document}
\title{Chiral scatterers designed by Bayesian optimization}

\author{Philipp Gutsche$^{1,2}$, Philipp-Immanuel Schneider$^3$,
Sven Burger$^{2,3}$ and Manuel Nieto-Vesperinas$^4$}

\address{
	$^1$Freie Universit\"at Berlin, Mathematics Institute, Arnimallee 6, 14195 Berlin, Germany \\
	$^2$Zuse Institute Berlin, Computational Nano Optics, Takustr.~7, 14195 Berlin, Germany \\
	$^3$JCMwave GmbH, Bolivarallee 22, 14050 Berlin, Germany \\
	$^4$Instituto de Ciencia de Materiales de Madrid, Consejo Superior de Investigaciones Cient\'ificas,
		Campus de Cantoblanco, Madrid 28049, Spain
}

\ead{gutsche@zib.de}

\begin{abstract}
The helicity or chirality of scattered light is strongly linked to the dual symmetry of the scatterer.
The latter depends on chiral materials or on scatterers which are not superimposable
with their mirror image. This inherently yields asymmetric structures of various shapes with many degrees
of freedom. In order to explore these high dimensional parameter spaces, numerical simulations and especially
optimization strategies are a valueable tool. Here, we optimize the emission of chiral line sources in two-dimensional dimer
setups using Bayesian optimization. We deduce relevant objective functions from recent theoretical findings
for chiral electromagnetic fields and employ rigorous simulations of Maxwell's equations.
\end{abstract}

\section{Introduction}

Several degrees of freedom of light including frequency, amplitude and polarization are controlled and tailored
for applications ranging from optical communication to molecular spectroscopy. Recently, more complex light fields
such as occuring in spin-orbit interactions \cite{bliokh2017} have been investigated.
Due to its close relation to these
phenomena, the helicity of electromagnetic fields is becoming an important quantity in the analysis of nano-optical
devices \cite{nieto2017a}.

Furthermore, nano-antennas have been proven a flexible tool for the enhancement of the emission of electromagnetic
sources \cite{abass2016}. In order to increase the contrast of signals from handed sources which possess opposing
helicities, it has been shown in the framework of dual symmetry \cite{fernandez2014} that asymmetric scatterers
or metamaterials are required in the nano-optical regime where most materials are non-magnetic.

Often, numerical simulations are carried out for obtaining suitable structures. However, due to the required
asymmetry, the underlying parameter spaces are mostly high-dimensional and naive optimization strategy fail
to provide fast and reliable results. In addition, fabrication requirements of the devices and/or
limitations in the illumination techniques need to be taken into account. That is why, sophisticated optimization
algorithms are required \cite{schneider2017}.

Here, we summarize recent descriptions of the helicity and, in its time-harmonic form, the chirality of
light: In section \ref{sec:helicity}, we describe the Helicity Optical Theorem and a Chirality Conservation Law.
Precedingly, the stochastical Bayesian optimization \cite{shahriari2016} is briefly described.
Finally, in section \ref{sec:opti}, we will apply both, the theory on helicity and the described optimization strategy, to the control
of two-dimensional line sources embedded in magneto-dielectric dimers.

\section{Description of optically chiral fields}
\label{sec:helicity}

In standard electrodynamics, dipolar fields are described by the electric $\pp$ and magnetic $\mm$
dipole moments. The electric and magnetic fields are up to a prefactor of the wave impedance
$Z = \sqrt{\mue/\epsi}$ dual to one another \cite{jackson1998}, where $\epsi$ is the permittivity
and $\mue$ is the permeability of the medium in which the source is embedded.

Recently, interest in chiral optical fields has been intensified due to the possibility to enhance
the near-field response of chiral molecules. By boosting the optical chirality density locally,
extinction measurements gain sensitivity \cite{tang2010}. Additionally, the chirality or helicity
of light serves as a tool for fundamental understanding of light-matter interactions \cite{fernandez2014}.

Most optical effects can be described by the electric dipole response. In order to observe
chiral effects, the quadrupole contribution of the magnetic dipole response needs to be taken
into account. Especially, chiral dipoles of well-defined helicity are of interest in this field.
Instead of the dipole basis of electric and magnetic dipoles, the chiral dipole basis of positive and negative
handedness with $\pp = \pm i\mm$ has been proven suitable for the description of chiral phenomena
\cite{zambrana2016}.

\subsection{Helicity Optical Theorem (HOT)}

The response of general dipolar sources $(\pp,\mm)$ to an incident electric and magnetic field
($\Ethinc,\Hthinc$) is described by the optical theorem:
\begin{align}
 \frac{\omega}{2} \imag{\pp \cdot \Ethinc^* + \mm \cdot \Bthinc^*} =
		\mathcal{W}^\text{abs} + \frac{c k^4}{3 n} \left( \frac{1}{\epsi} |\pp|^2 + \mue |\mm|^2 \right).
\end{align}
The left-hand side describes the extinction of energy, $\mathcal{W}^\text{abs}$ is the energy
absorption and the remaining part is the scattered energy.

In addition to the analysis of energy quantities, a Helicity Optical Theorem has been
formulated \cite{nieto2017a}:
\begin{align}
	\frac{2\pi c}{\mue} \real{-\frac{1}{\epsi} \pp \cdot \Bthinc^* + \mue \mm \cdot \Ethinc^*} =
		\mathcal{W}^\text{abs}_\mathcal{H} + \frac{8 \pi c k^3}{3 \epsi} \imag{\pp \cdot \mm^*}.
\end{align}
Here, $\mathcal{W}^\text{abs}_\mathcal{H}$ is the absorption of helicity (or the conversion of chirality
as explained in the next section). The left-hand side is the extinction of chirality and it is obvious
that the scattered chirality $\imag{\pp \cdot \mm^*}$ is only non-zero if the electric and magnetic dipole
moments are phase-shifted. The chiral dipoles of well-defined helicity mentioned above scatter
the magnitude of $\pm|\pp|^2$ chirality.

This formalism is well suited for the study of extinction experiments with secondary dipolar sources. In the following,
we show an equivalent description in the time-harmonic context: the general chirality conservation law.
Proceedingly, we apply this law to the analysis of a primary dipolar source.

\subsection{Chirality Conservation Law (CCL)}

The interest in the quantity of optical chirality is manifested in its ability to be described
by a general conservation law. This conservation can be formulated for arbitrary, including bi-anisotropic, media.
The most general form for monochromatic fields read as \cite{gutsche2016}
\begin{align}
	2 i \omega (\chielth - \chimath) + \nabl \cdot \Sigmth = \frac{1}{4}
		\left\{ \Jth^* \cdot \rot{\Eth} + \rot{\Jth^*} \cdot \Eth \right\}.
	\label{eq:chCont}
\end{align}
Similarly to the electromagnetic energy, the chirality density is split into an electric
$\chielth = 1/8 \left\{ \Dth^* \cdot \rot{\Eth} + \Eth \cdot \rot{\Dth} \right\}$ and
magnetic part $\chimath = 1/8 \left\{ \Hth^* \cdot \rot{\Bth} + \Bth \cdot \rot{\Hth} \right\}$.
The flux density of optical chirality is $\Sigmth = 1/4 \left\{ \Eth \times \rot{\Hth} - \Hth^* \times \rot{\Eth} \right\}$.
The source term on the right-hand side of this continuity equation is of special interest for this study.

Using the current densities $\Jth$ for electric and magnetic dipoles, \eqref{eq:chCont} yields the HOT
in the context of extinction measurements.
For the analysis of primary sources, analogous to the case of energy, the conservation of chirality
describes the emission $\Xemi$, conversion $\Xconv$ and the far-field accessible chirality flux $\Xfar$.
These quantities are employed for the optimization of chiral effects in the preceeding study and are
for localized sources derived as:
\begin{align}
	\Xemi &= X_0 + \frac{1}{2} \omega \int \imag{\Jth^* \cdot \Bthsca} d^3r \label{eq:Xemi} \\
	\Xconv &= -2 \omega \int \imag{\chielth - \chimath} d^3r \\
	\Xfar &= \int_{S^2(r\to\infty)} \real{\Sigmth} d^2r = \Xemi - \Xconv.
\end{align}
The chirality emission $\Xemi$ consists of the bulk chirality emission $X_0$, which is the chirality emitted from
the dipole in absence of any interacting environment, e.g.~$X_0=0$ for achiral
sources. In addition, a term occurs in $\Xemi$ which depends on the current density $\Jth$, i.e.~in our case the dipole moments, and
the magnetic field $\Bthsca$ scattered by the object. The integration of the chirality flux in the far-field,
$r \to \infty$, yields the far-field accessible quantity $\Xfar$. The chirality conversion $\Xconv$
consists of a bulk and an interface term for piecewise-constant materials \cite{gutsche2016}.

\section{Bayesian optimization}

Local optimization is a fairly simple task with various well-established methods such as
gradient descent or Newton algorithms. However, in complex systems several optima might arise
and mostly only the global optimum is of interest. Bayesian optimization with Gaussian processes
tackles this problem and is applied in many fields such as robotics, experimental design and
environment monitoring.

Recently, it has also been applied to nano-optical devices \cite{schneider2017}. Based on this
study, we summarize the basic formalism underlying Bayesian optimization and
give details on our implementation in the following.

\subsection{Basic formalism}
The aim of an optimization is to minimize an objective function $f$ depending on the parameters given by $\xx$:
\begin{align}
	\widetilde{\xx} = \underset{\xx \in \mathcal{X}}{\operatorname{arg min}} f(\xx),
\end{align}
where $\widetilde{\xx}$ are the parameters of the global minimum of $f$ in the $d$-dimensional
design space $\mathcal{X} \subset \mathbb{R}^d$. The design space might be a $d$-dimensional hypercube or
a general constraint parameter space.

In each search step Bayesian optimization (BO) takes all previously obtained evaluations of $f$ into account
to determine promising parameters and is thus able to reach the global minimum with relatively few iterations.
BO is based on a statistical model. Due to their flexibility and tractability, Gaussian processes
have been proven to be a suitable basis for this model. Here, every evaluation is associated
with a normally distributed random variable.

A Gaussian process is defined by a mean function $\mue: \mathcal{X} \to \mathbb{R}$ and a kernel function
$k: \mathcal{X} \times \mathcal{X} \to \mathbb{R}$ which is positive definite and describes the covariance of the
process. The probablity of the multivariate Gaussian random variable $\vec{Y} = [f(\xx_1), ...,f(\xx_N)]^T$
is given by
\begin{align}
	P(\vec{Y}) = \frac{1}{(2\pi)^{N/2} |\Sigm|^{1/2}}
		\exp \left[-\frac{1}{2} (\vec{Y} - \vec{\mue})^T \Sigm^{-1} (\vec{Y} - \vec{\mue}) \right],
\end{align}
where $\vec{\mue} = [\mue(\xx_1), ..., \mue(\xx_N)]^T$ are the mean values and $\Sigm = [k(\xx_i,\xx_j)]_{i,j}$
is the covariance matrix.

Based on already known values of the objective function, it is possible to make statistical predictions about
not yet evaluated points in the parameter space by Gaussian process regression. The determination of the next
evaluation point is based on the predicted Gaussian distribution of the function value (mean and standard deviation).
This information is used in the determination of
the next evaluation point $\xx_{N+1}$ by an acquisition function
$\alpha: \mathcal{X} \times \mathbb{R} \to \mathbb{R}$:
\begin{align}
	\xx_{N+1} = \underset{\xx \in \mathcal{X}}{\operatorname{arg max}} ~ \alpha(\xx, y_\text{min}),
\end{align}
where $y_\text{min} = \min(f(\xx_1), ..., f(\xx_N))$ is the currently obtained global minimum.

\subsection{Implementation details}
Our implementation of BO is based on the Matern-5/2 kernel
\begin{align}
	k(\xx,\xx') = \sigma^2 \left( 1 + \sqrt{5 r^2(\xx,\xx')} + \frac{5}{3} r^2(\xx,\xx') \right)
		\exp \left( -\sqrt{5 r^2(\xx,\xx')} \right)
		\\ \nonumber
		\text{with~} r^2 = \sum_{i=1}^d (x_i - x_i')^2 / l_d^2,
\end{align}
where $\sigma$ is the standard deviation and $l_1,...,l_d$ describe the length scales of the parameters.
As noted in \cite{schneider2017} this kernel leads to slightly faster optimizations than the frequently used squared exponential kernel.

The hyper-parameters of the Gaussian process $\omega = (\sigma,l_1,...,l_d)$ are unkown. Since they
are at the core of the BO optimization strategy, they should be chosen in an optimal fashion.
This optimization is a computationally very expensive part of BO.
The employed strategy to reduce the computational overhead of the hyper-parameter optimization is presented in \cite{garcia2017}.

For the acquisition function $\alpha$, we use the expected improvement
\begin{align}
	\alpha(\xx, y_\text{min}) = \mathbb{E}[\max_{\xx}(0,y_\text{min}-f(\xx))],
\end{align}
where $f(\xx)$ is the statistical prediction (Gaussian distribution) of the unknown objective function at the position $\xx$.
That is, we take the next sample at the point in the parameter space where the expectation value of the one-sided difference
$\max_{\xx}(0,y_\text{min} −f(\xx))$ between the currently known minimum $y_\text{min}$ and the Gaussian distribution of $f(\xx)$ is maximized.

The BO employed in this study is implemented in \textit{python}. The Gaussian processes are based
on the module \textit{gptools}. Further details may be found in \cite{schneider2017} and \cite{garcia2017}.

\section{Automatic design of optically chiral dimers for line sources}
\label{sec:opti}

In this study, we present the application of BO to a magneto-dielectric dimer showing optically chiral behaviour.
Different aims and accordingly different objective functions are employed for the optimization of
dual symmetry, i.e.~conservation of helicity, as well as the selective emission enhancement for
one handedness of dipolar chiral sources. Finally, a scatterer is designed preserving one handedness and
changing the opposite handed source into an achiral field in the far-field.

\subsection{Setup and parametrization}

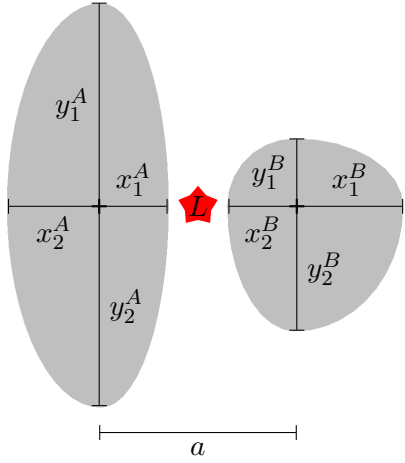
\begin{figure}[h]
	\begin{center}
  \begin{tikzpicture}
    \node at (0,0) {
      \includegraphics[width=0.35\textwidth,trim=250 10 250 10,clip]{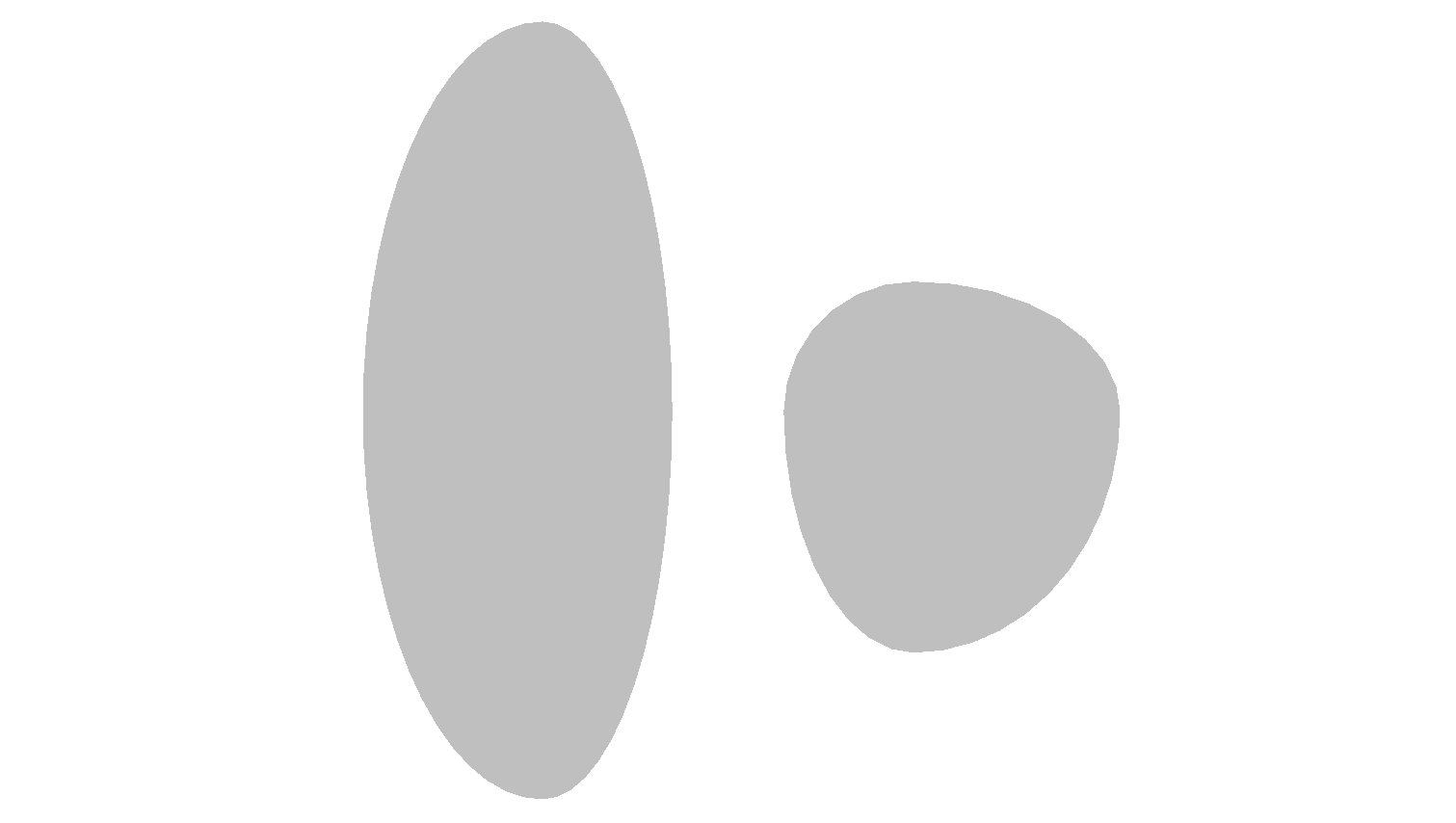}
    };  
    \node[red,fill=red,star] (0,0) [radius=0.25] {}; \node at (0,0) {$L$};
    
    \node (CA) at (-1.3,0) {}; 
      \node (TA) at (-1.3,2.7) {}; 
      \node (BA) at (-1.3,-2.65) {}; 
      \node (LA) at (-2.5,0.0) {}; 
      \node (RA) at (-0.4,0.0) {}; 
    \draw[|-|] (CA.center) -- (TA.center) node[pos=0.5, left] {$y_1^A$};
    \draw[|-|] (CA.center) -- (BA.center) node[pos=0.5, right] {$y_2^A$};
    \draw[|-|] (CA.center) -- (RA.center) node[pos=0.5, above] {$x_1^A$};
    \draw[|-|] (CA.center) -- (LA.center) node[pos=0.5, below] {$x_2^A$};
    
    \node (CB) at (1.3,0) {}; 
      \node (TB) at (1.3,0.9) {}; 
      \node (BB) at (1.3,-1.65) {}; 
      \node (RB) at (2.7,0.0) {}; 
      \node (LB) at (0.4,0.0) {}; 
    \draw[|-|] (CB.center) -- (TB.center) node[pos=0.5, left] {$y_1^B$};
    \draw[|-|] (CB.center) -- (BB.center) node[pos=0.5, right] {$y_2^B$};
    \draw[|-|] (CB.center) -- (RB.center) node[pos=0.5, above] {$x_1^B$};
    \draw[|-|] (CB.center) -- (LB.center) node[pos=0.5, below] {$x_2^B$};

		\node (CAA) at (-1.3,-3) {};	
		\node (CBB) at (1.3,-3) {};	
    \draw[|-|] (CAA.center) -- (CBB.center) node[pos=0.5, below] {$a$};
  \end{tikzpicture}
	\end{center}
	\caption{
		Parametrization of dimer consisting of constitutens $A$ and $B$ each with four different radii
		$x_1, y_1, x_2$ and $y_2$ and separated by distance $a$. Line Source $L$ is placed in the center of the dimer.
		\label{fig:geo}
	}
\end{figure}

The scatterer analyzed in the following is a dimer consisting
of constitutents $A$ and $B$ (Fig.~\ref{fig:geo}). The centre points of $A$ and $B$ are separated by the
distance $a$. Each constitutent is parametrized by four radii, respectively. This results in a shape
given by $(x_i \cos \beta, y_j \sin \beta)^T \subset \mathbb{R}^2$, where $x_i$ and $y_j$ are chosen
according to the quadrant and $\beta$ is the corresponding angle. For example, in the second quadrant:
$i = 2, j = 1, \beta \in [\pi/2, \pi]$.

The material of the dimer has a high refractive index of $n=4$.
Accordingly, magnetodielectric effects with high magnetic resonances can be observed, although the material
is purely electric with relative permittivity $\epsi_r = n^2$ and relative permeability $\mue_r=1$ \cite{krasnok2016}.
This is crucial for the observation of optically chiral behaviour, since for this application spectrally matched electric
and magnetic resonances of similar strength are required.

In the centre of the dimer, we place a line source emitting at wavelength $\lambda$.
Chiral line sources with $\pp = \pm i\mm$ are analyzed. In order to observe chiral effects despite the underlying
$z$-invariance of the setup, we use $\pp \propto (0,1,1)^T$.
The standard Purcell factor of this two-dimensional source differs only in its prefactor from the well-known
formula for a three-dimensional dipole \cite{novotny2006}:
\begin{align}
	F_P = 1 + \frac{1}{2W_0} \left\{ \frac{1}{\omega\mue} \imag{\pp^* \cdot \Ethsca} + k \imag{\mm^* \cdot \Hthsca} \right\},
\end{align}
where $W_0$ is the energy radiated by the dipole alone, in absence of any interacting environment.
The corresponding chirality emission in \eqref{eq:Xemi} for a magnetic and electric dipolar sources is \cite{nieto2017a}
\begin{align}
	\Xemi = X_0 + \real{\pp^* \cdot \Hthsca} - \frac{1}{Z} \real{\mm^* \cdot \Ethsca}.
\end{align}
The electromagnetic response of the devices are computed from rigorous solutions of Maxwell's equations.
These are obtained with the Finite Element Method implemented in the commercial Maxwell solver \textit{JCMsuite}
\cite{pomplun2007}.

\subsection{Optimization of dual elliptical dimers}

In order to present the optimization procedure in a low-dimensional space, we fix the emission wavelength
$\lambda = 950\text{nm}$ and the distance $a = 200 \text{nm}$.
Furthermore, we set $x_1 = x_2$ and $y_1 = y_2$ for both constitutents $A$ and $B$ of the dimer.
In addition, the dimer is mirror symmetric, i.e.~$x_A = x_B$ and $y_A = y_B$. This leaves two free parameters,
$x$ and $y$ for the optimization. Accordingly, $A$ and $B$ are ellipses and it has been shown that
ellipsoids are suitable for matching electric and magnetic resonances \cite{luk2015}.

The objective function is chosen as the far-field helicity
\begin{align}
	h = \frac{\Xfar}{k \Wfar},
	\label{eq:hel}
\end{align}
where $\Xfar$ is the chirality and $\Wfar$ is the energy measured in the far-field, respectively.
The absolute value of $|h| \leq 1$ is limited to unity.
Using a negative chiral source, the helicity $h=-1$ describes so-called \textit{dual} behaviour \cite{zambrana2016} of the dimer:
The chirality observed in the far-field is well-defined and has the same value as the chirality of the emitter.

We use BO to minimize $h$ with respect to the ellipse radii $x \in [20, 80]\text{nm}$ and $y \in [20,210]\text{nm}$.
For 16 parallel evaluations of the objective function, the BO requires 122 evaluations in total to obtain
a global minimum of approximately $-0.956$. This means that only less than $4.4\%$ of the emitted chirality
is not preserved.
The optimal parameters are $x \approx 68.66\text{nm}$ and $y \approx 150.64\text{nm}$.

\begin{figure}[h]
	\begin{minipage}{0.45\textwidth}
		\includegraphics[width=\textwidth]{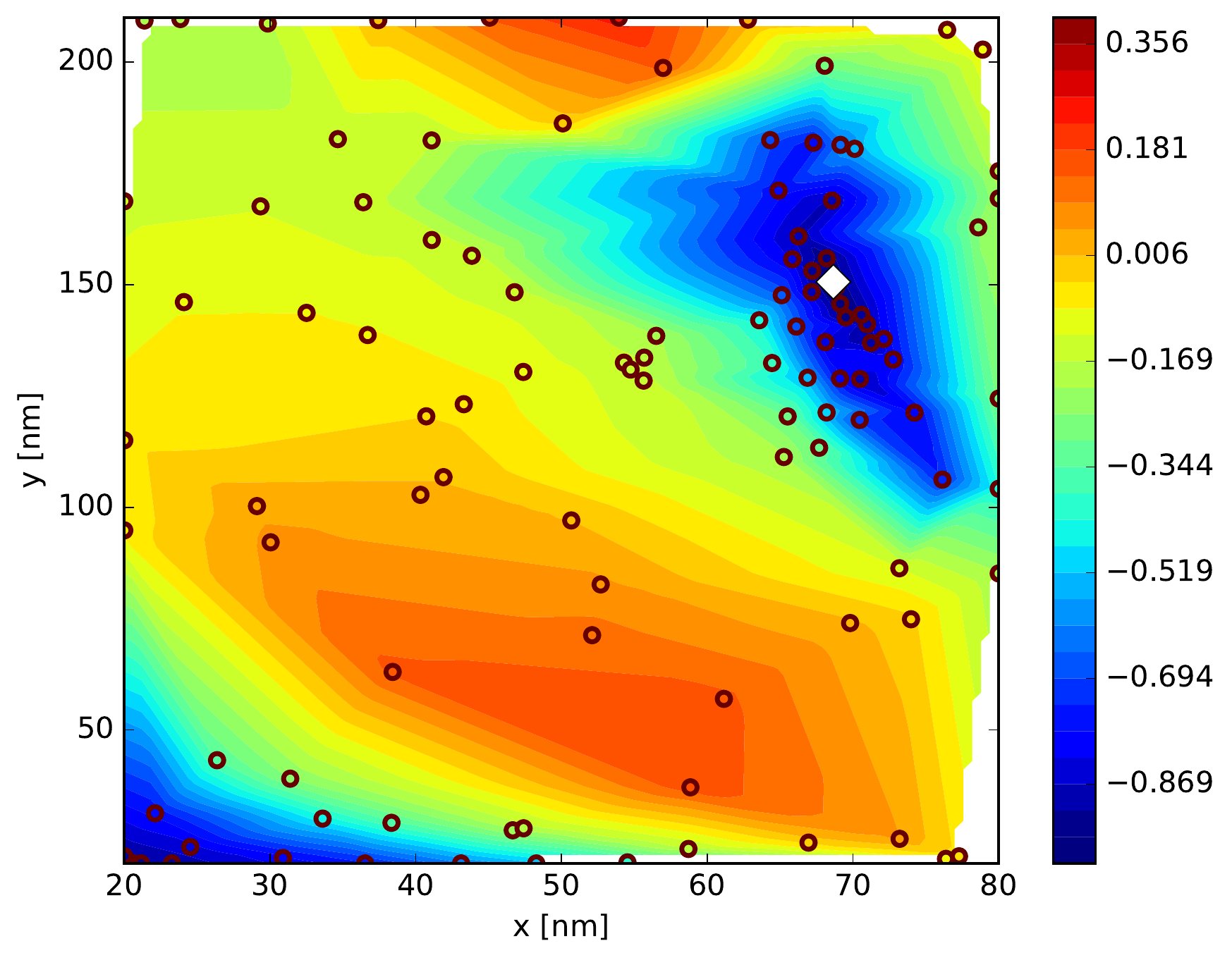}
		\caption{
			Parameters obtained by BO. Colormap shows Gaussian regression.
			\label{fig:ellip:scan}
		}
	\end{minipage}
	\hfill 
	\begin{minipage}{0.45\textwidth}
		\includegraphics[width=\textwidth]{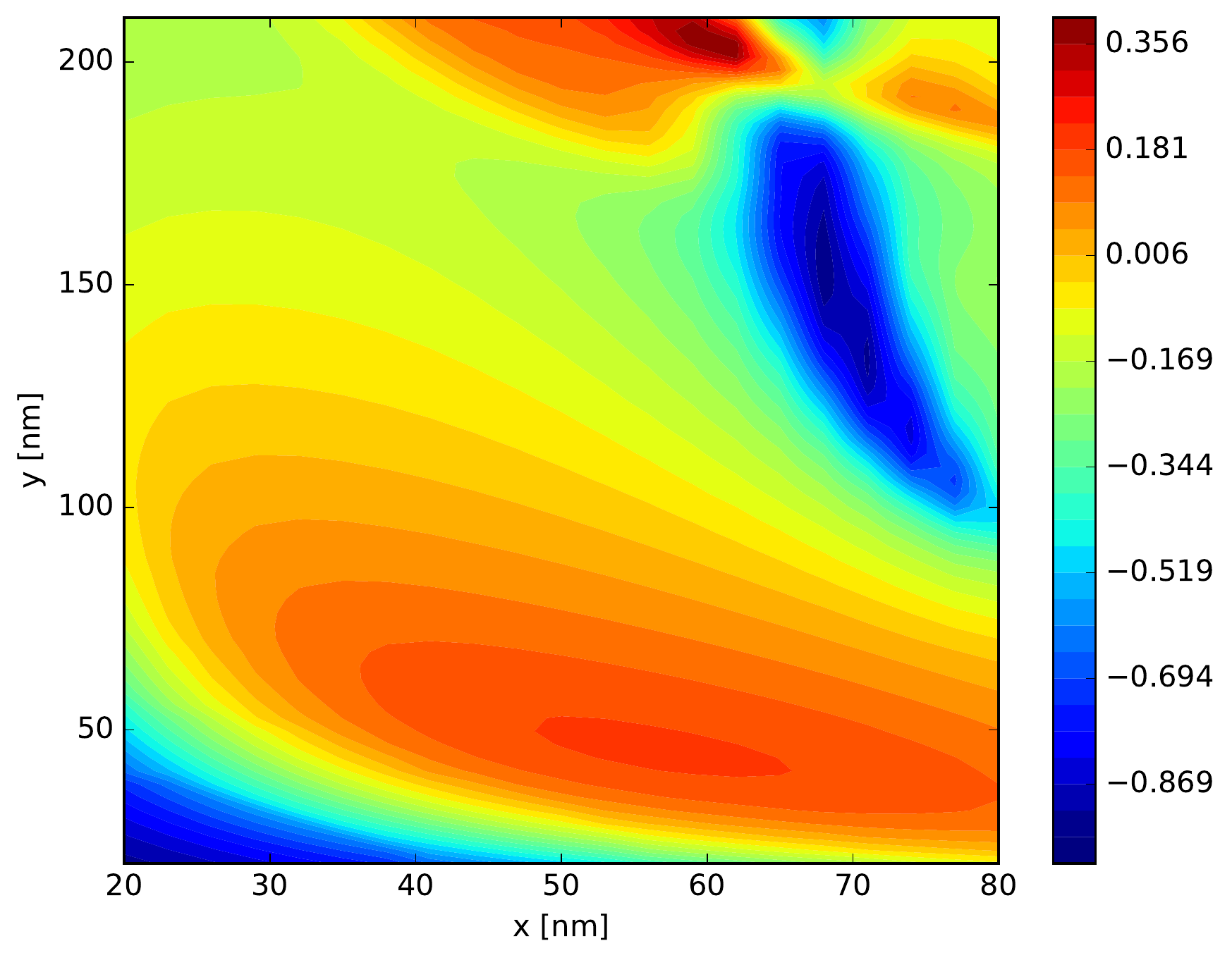}
		\caption{
			Full parameter space. Colormap shows the objective function.
			\label{fig:ellip:full}
		}
	\end{minipage} 
\end{figure}

As shown in Fig.~\ref{fig:ellip:scan}, many different values of $h$ are obtained in the optimization process.
The acquired parameters points and their respective objective values are depicted by circles.
In addition, Gaussian regression allows for predicting $h$ in the full parameter space and this is shown by the colormap.

In this low dimensional space it is possible for reasonable computational costs to scan the full parameter space
which is shown in Fig.~\ref{fig:ellip:full}. In the chosen domain there is only one minimum prominent alongside
a larger minimum at the lower corner of the domain. The prediction from BO
is very accurate, especially in the surrounding of the global minimum.

\subsection{Optimization of optically chiral dimers}

In the following, we allow optimization of all eight radii and the dimer distance (Fig.~\ref{fig:geo})
resulting in nine
free parameters for the BO. Again, we fix the emission wavelength $\lambda=950\text{nm}$.
Assuming non-dispersive media, varying the frequency of the emitter requires only scaling of all geometrical
parameters of the optimized structure.

The $x$-parameter space is limited to $[20,80]\text{nm}$ and
the $y$ parameter space is restricted to $[20,210]\text{nm}$. The distance $a$ is limited to $[40,350]\text{nm}$.
Furthermore, the following constraint in the parameter space is employed in order to ensure a minimal distance of the
source and the scatterer: $\max(x_1^A, x_2^B) + 20nm - 0.5a < 0$.

\begin{center}
	\begin{figure}[h]
		\begin{minipage}{0.35\textwidth}
			\vspace{8ex}
			\begin{center}
  			\begin{tikzpicture}
    			\node at (0,0) {\includegraphics[width=\textwidth]{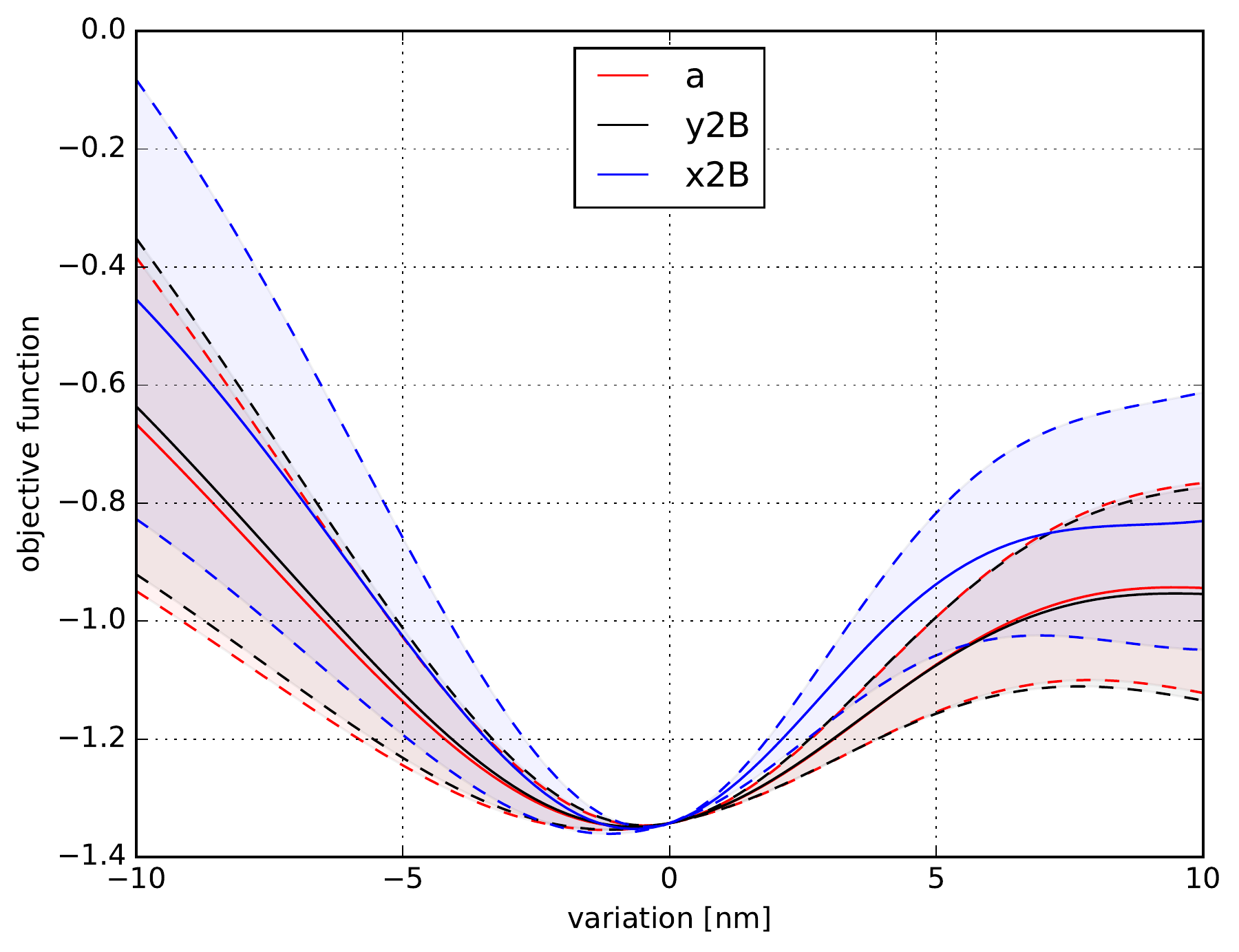}};  
    		\end{tikzpicture}
			\end{center}
			\caption{
				Optimization of $f_{g_P}$. Most sensitive parameters.
				\label{fig:obj0}
			}
		\end{minipage}
		\hfill
		\begin{minipage}{0.6\textwidth}
			\begin{center}
			\begin{tabular}{c||c|c|c|c}
				obj.~fct. & $F_P^-$ & $F_P^+$ & $g_P$ & opt.~geom.
				\\ \hline
				$f_{g_P}$ & 8.38 & 1.65 & -1.34 &
					\raisebox{-8ex}{
    			\includegraphics[width=0.25\textwidth]{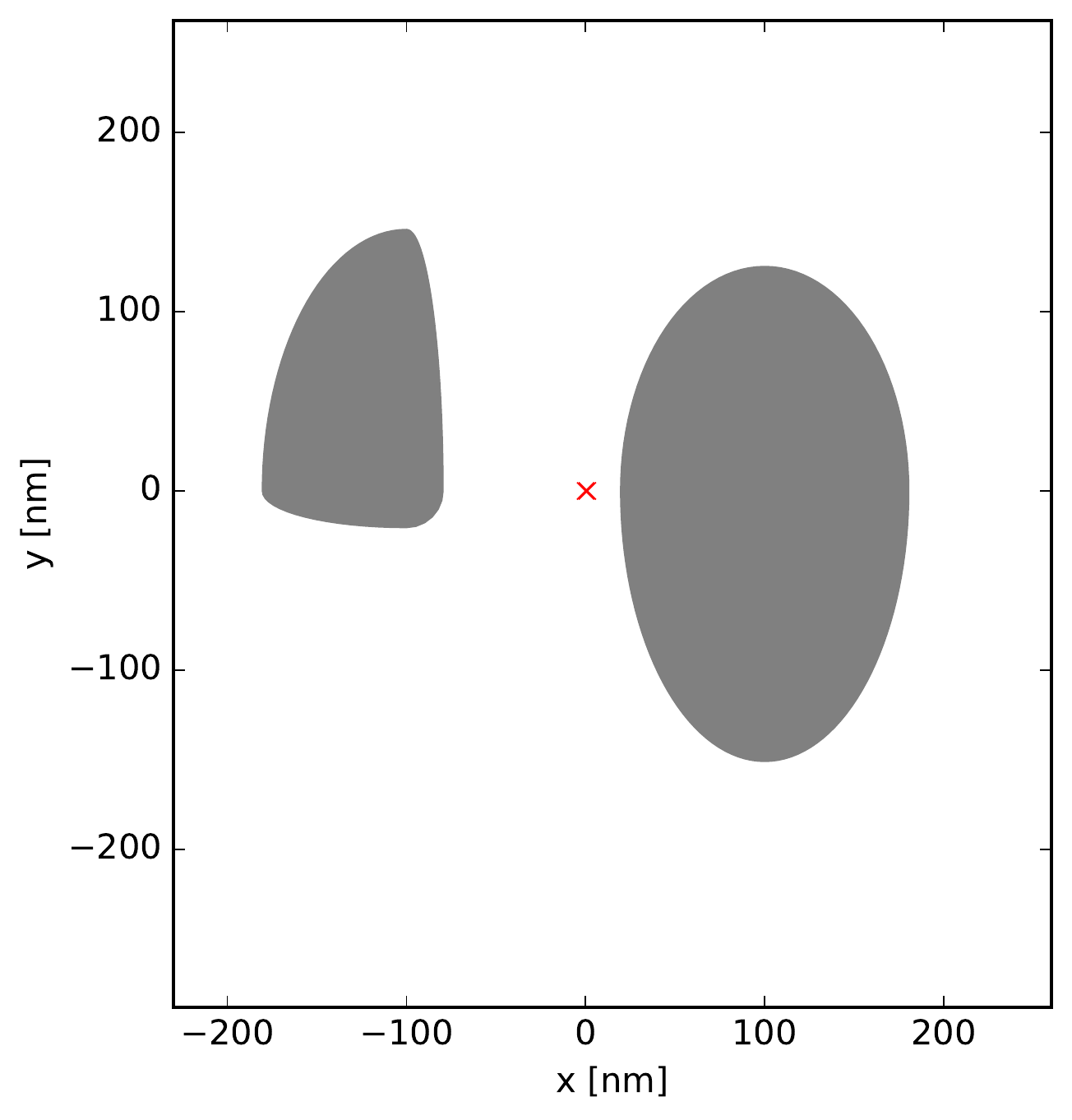}
					}
				\\ \hline
				$\widetilde{f}_{g_P}$ & 9.56 & 2.09 & -1.28 &
					\raisebox{-8ex}{
    			\includegraphics[width=0.25\textwidth]{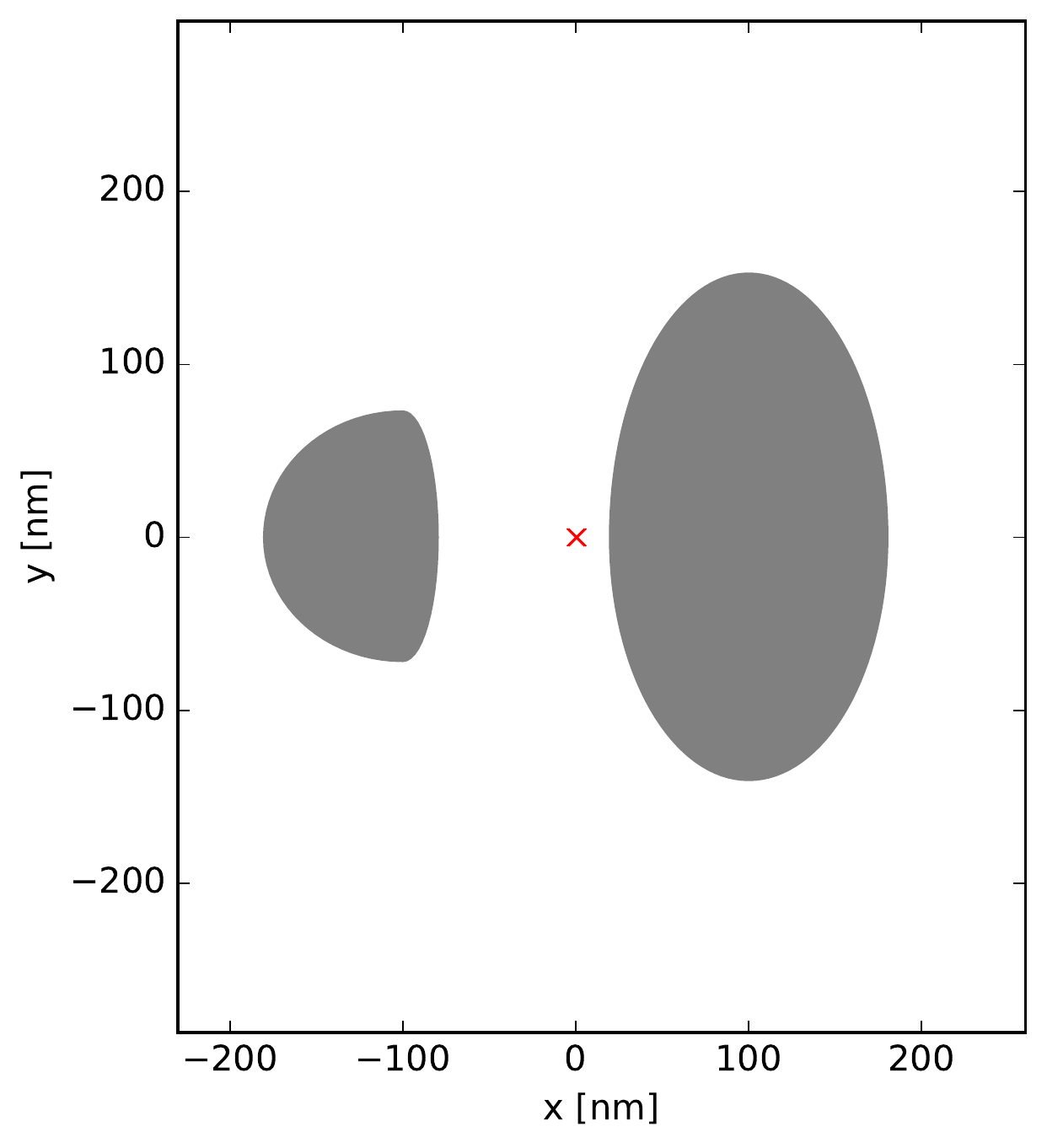}
					}
			\end{tabular}
			\end{center}
			\caption{
				Optimization of $f_{g_P}$ and $\widetilde{f}_{g_P}$.
				Purcell factors, their relative difference and optimal geometries.
				\label{fig:obj1}
			}
		\end{minipage}
	\end{figure}
\end{center}

For non-mirror-symmetric dimers, it is possible to observe optically chiral effects in the sense that
emission of positive and negative chiral sources differ. According to the experimental setup, different
quantities might be of interest for the optimization. Here, we optimize for four different objective functions:
\begin{itemize}
	\item the relative difference of Purcell factors of positive and negative chiral sources
		\begin{align}
			f_{g_P}(\xx) = g_P = 2 \frac{F_P^+ - F_P^-}{F_P^+ + F_P^-}
			,~~~ f_{g_P} \in [-2, 2]
			\label{eq:fgp}
		\end{align}
	\item the relative difference of Purcell factors ensuring a large $F_P^-$
		\begin{align}
			\widetilde{f}_{g_P}(\xx) = \left\{ \frac{1}{4} g_P + \frac{1}{2} \right\} + \frac{1}{F_P^- + 1}
			,~~~ \widetilde{f}_{g_P} \in (0, 2)
			\label{eq:tfgp}
		\end{align}
	\item the relative difference of scattered helicity in the far-field
		\begin{align}
			f_{g_\text{far}}(\xx) = g_\text{far} = |h^+| - |h^-|
			,~~~ f_{g_\text{far}} \in [-1, 1]
			\label{eq:fgfar}
		\end{align}
	\item the relative difference of scattered helicity ensuring a large $F_P^-$
		\begin{align}
			\widetilde{f}_{g_\text{far}}(\xx) = \left\{ \frac{1}{2} g_\text{far} + 0.5 \right\} + \frac{1}{F_P^- + 1}
			,~~~ \widetilde{f}_{g_\text{far}} \in (0, 2)
			\label{eq:tfgfar}
		\end{align}
\end{itemize}
Fig.~\ref{fig:obj0} shows the result of the optimization of all nine parameters for the relative difference of
Purcell factors $f_{g_P}$. The sensitivity
of the objective value with respect to the three most sensitive parameters $a$, $y_2^B$ and $x_2^B$ is given.
On the $x$-axis the variation
of the parameters in nm and on the $y$-axis the predicted change of the objective function is
shown. These predictions are derived from Gaussian regression and their standard deviation is depicted
with dashed lines.

The limits in the parameter space are analogous to the optimization of the dual dimer in the previous section.
The detailed geometric parameters are given in Tab.~\ref{tab:fgp}.
Five out of nine parameters are at the upper or lower limit of the parameter
hypercube. A larger parameter space might therefore be beneficial for further optimization, however, larger
violation of the dual symmetry is expected.

A minimal $g_P$-factor of -1.34 is achieved by a highly achiral scatterer. Since the emission of negative chiral
sources is larger for a negative $g_P$-factor, the constitutent $B$ is as close as possible in the limited
parameter space to the source which emits mostly to the right. However, a minimized $g_P$ could also be achieved
by purely reducing $F_P^+$. That is why, we change to the objective function to $\widetilde{f}_{g_P}$ which
ensures a large $F_P^-$.

As shown in Fig.~\ref{fig:obj1}, the adjusted objective functions results in an increase of more than 10\% in
the Purcell factor of the negative chiral source $F_P^-$. On the other hand, $g_P$ is only decreased by less
than 5\%. These changes in the objective values are rather small. However, it shows that the objective function has to
be chosen carefully to obtain the desired behaviour of the device. This will be more important in the next example.

\begin{center}
	\begin{figure}[hb]
		\begin{minipage}{0.35\textwidth}
			\vspace{21ex}
			\begin{center}
  			\begin{tikzpicture}
    			\node at (0,0) {\includegraphics[width=\textwidth]{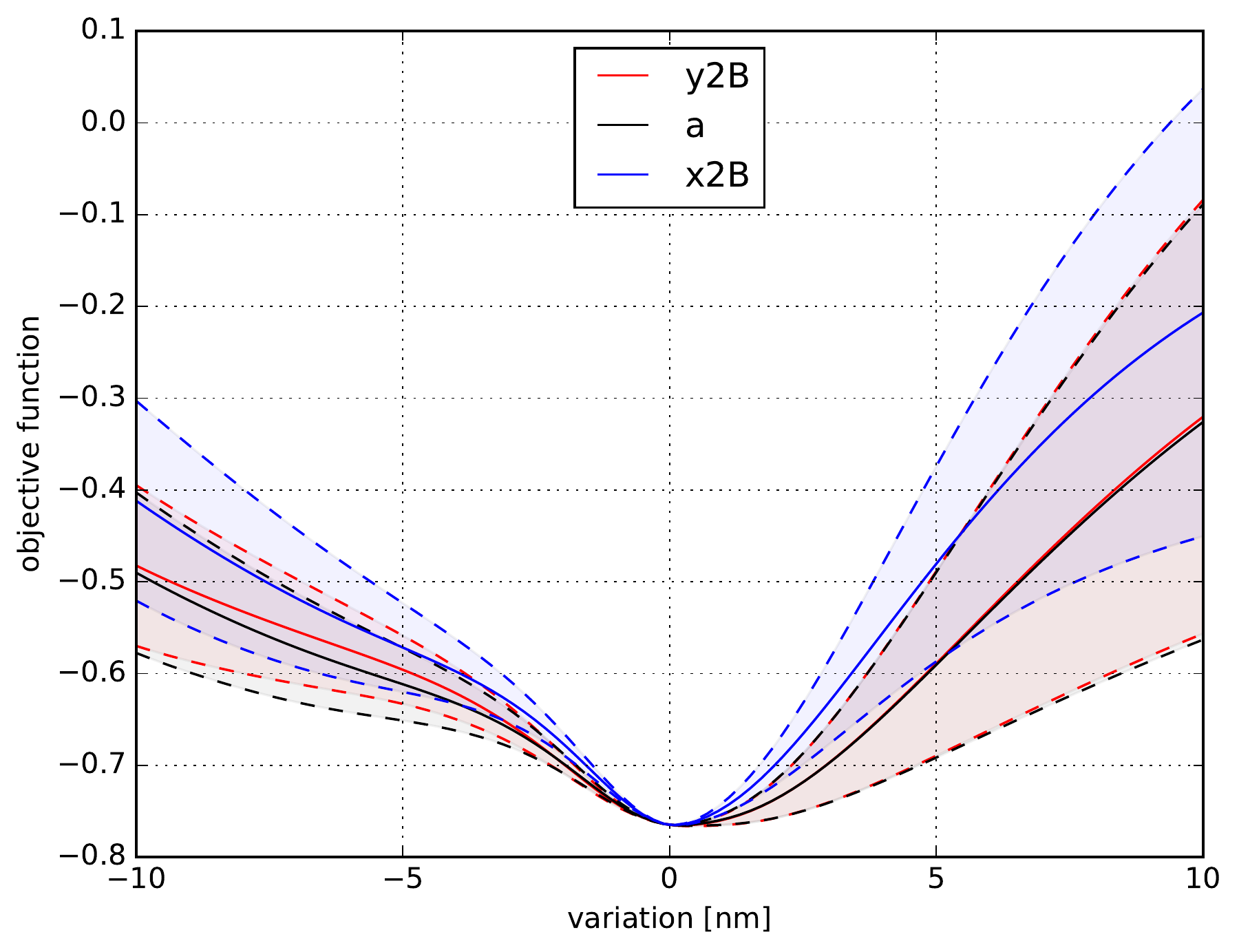}};  
    		\end{tikzpicture}
			\end{center}
			\caption{
				Optimization of $f_{g_\text{far}}$. Most sensitive parameters.
				\label{fig:obj2}
			}
		\end{minipage}
		\hfill
		\begin{minipage}{0.6\textwidth}
			\vspace{7ex}
			\begin{tabular}{c||c|c|c|c|c}
				obj.~fct. & $h^+$ & $F_P^-$ & $F_P^+$ & $g_\text{far}$ & opt.~geom.
				\\ \hline
				$f_{g_\text{far}}$ & 5.3e-6 & 1.45 & 0.88 & -0.77 &
					\raisebox{-8ex}{
    			\includegraphics[width=0.25\textwidth]{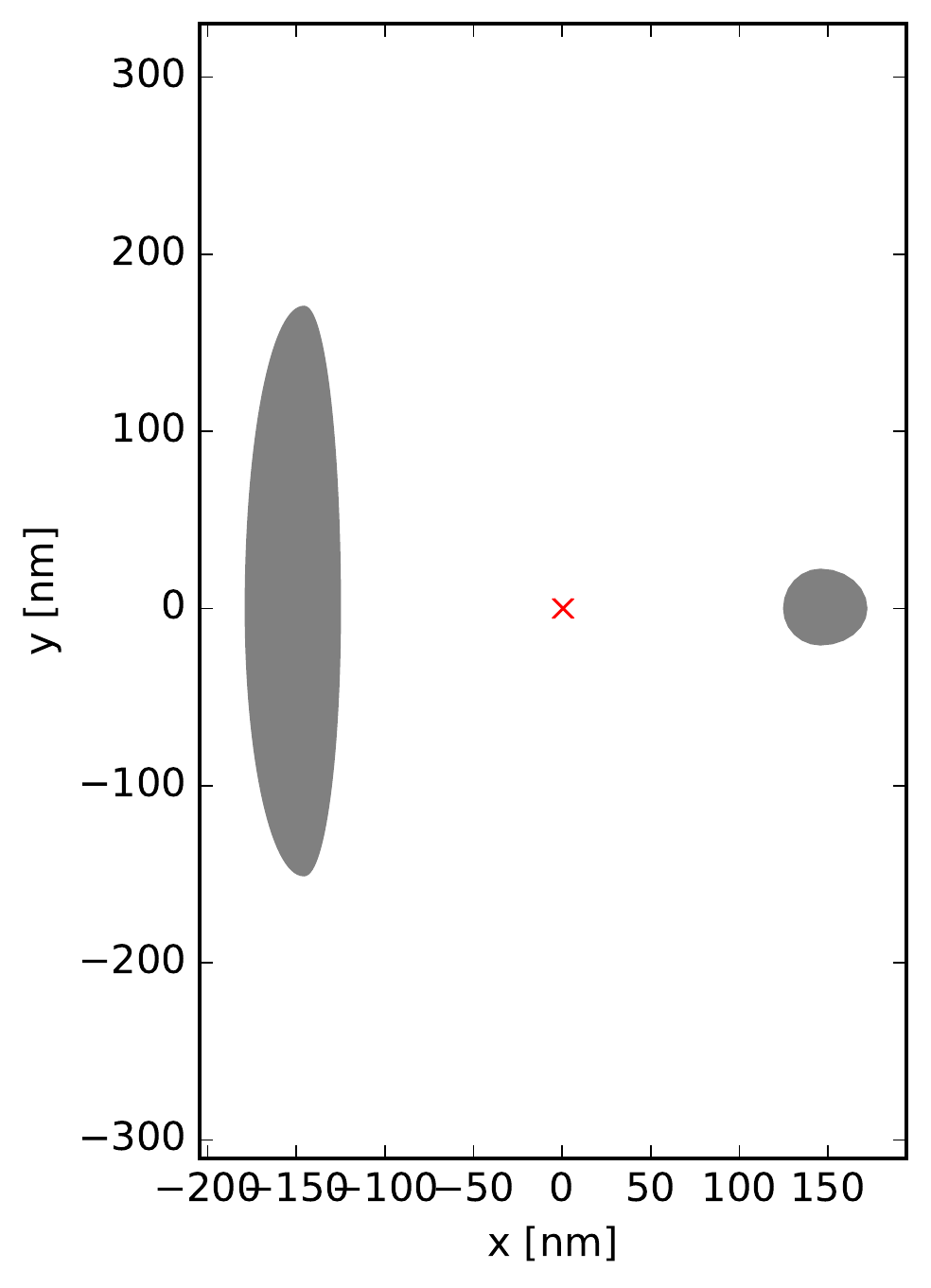}
					}
				\\ \hline
				$\widetilde{f}_{g_\text{far}}$ & -5.5e-3 & 3.57 & 1.66 & -0.68 &
					\raisebox{-8ex}{
    			\includegraphics[width=0.25\textwidth]{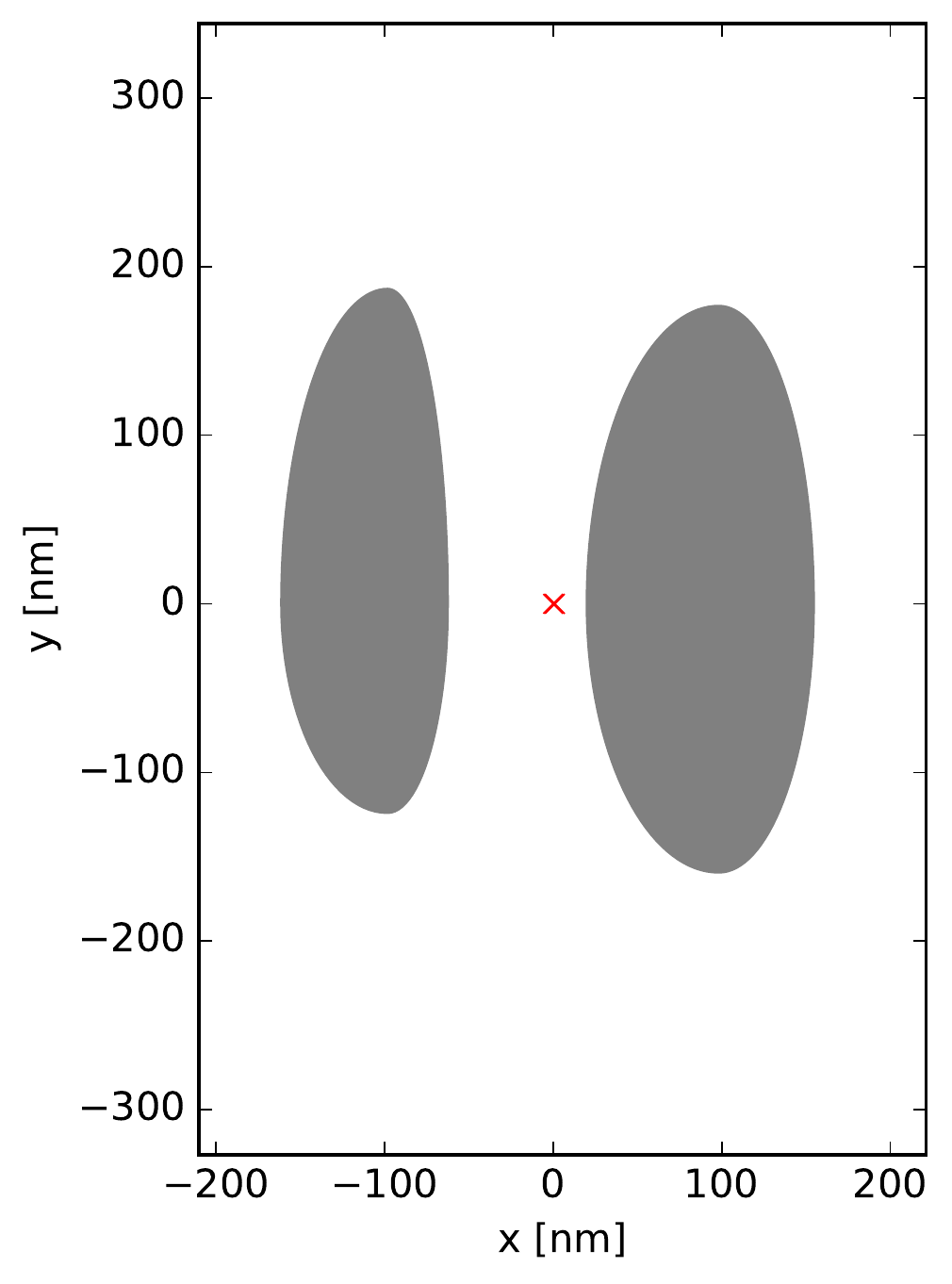}
					}
			\end{tabular}
			\caption{
				Optimization of $f_{g_\text{far}}$ and $\widetilde{f}_{g_\text{far}}$.
				Far-field chiralities, their rel.~difference and optimal geometries.
				\label{fig:obj3}
			}
		\end{minipage}
	\end{figure}
\end{center}

In addition to the standard Purcell factors, it provides new insight to analyze the helicity or
chirality of light. Namely, the far-field helicity \eqref{eq:hel} is of interest in the analysis
of e.g.~chiral molecules. Note that the investigated structure is lossless and accordingly
$\Wfar = \Wemi$, i.e.~the Purcell factor $F_P$ is experimentally accessible in the far-field.

Minimizing the objective function $f_{g_\text{far}}$ yields a structure which preserves helicity
for negative chiral sources and yields achiral emission in the far-field for positive chiral sources.
This could be of interest in order to distinguish both handedness with high accuracy.
Detailed geometric parameters of the optimal geometry are given in Tab.~\ref{tab:fgfar}.

As shown in Fig.~\ref{fig:obj2} and Fig.~\ref{fig:obj3}, the BO gives an optimal geometry which basically
turns the chiral emission of a positive chiral source into achiral fields in the far-field: the
helicity is below 1\%. On the other hand, chirality of negative chiral sources is dominantly preserved
and less than 23\% of positive chirality is observed in the far-field. Accordingly, $g_\text{far}$
of the optimized geometry is well suited to distinguish positive from negative chiral sources by analyzing the 
chirality observable in the far-field.

However, this comes with the disadvantage of relatively low emission enhancement of $F_P^- = 1.45$.
Consequently, we change the objective function from optimizing purely $g_\text{far}$ to
$\widetilde{f}_{g_\text{far}}$
which takes an increase in the Purcell factor into account. Fig.~\ref{fig:obj3} shows that the optimal
device is very different from the previous result. Instead of minimizing chirality conversion by
decreasing the area of the dimer constitutent $B$ to a minimum, the two parts of the dimer are of equal
size but different shape.

This approach increases the Purcell factor of the negative chiral source by a factor of nearly 2.5.
Nevertheless, $g_\text{far}$ is decreased to -0.68, still yielding basically achiral far-field behaviour
for a positive chiral source but showing higher chirality conversion for negative chiral sources.
Since the ratio of $F_P^-$ and $F_P^+$ is higher for the second optimization, we expect this
device having better experimental performance.

In summary, we have shown for standard energy quantities such as the difference of Purcell factors of chiral
sources as well as for novel quantities such as the helicity of light that the objective function of the optimization
has to be chosen carefully. In the best case, a single objective function is able to take several design
goals such as high distinction of positive and negative chiral sources as well as high emission enhancement
into account. The choice of different objective function results in geometrically much different
devices.

\section{Conclusion}
We used Bayesian Optimization to optimize the design of two-dimensional dimer structures.
Quantities of interest were derived from a general theory of the helicity or chirality
of light, namely, the recently introduced Helicity Optical Theorem and Chirality
Conservation Law. Even in high-dimensional parameter spaces with nine free parameters,
BO is a computationally feasible approach to minimize the number of objective function 
evaluations and obtain the global minimum in a given parameter space.

We expect that this numerical technique together with the rigorous physical description 
of optically chiral phenomena, provides further insight into devices for the analysis of chiral
phenomena and
could give guidelines for experimental realization of measurements of the chirality of
light.

\pagebreak
\ack{
We acknowledge support by Freie Universit\"at Berlin through the Dahlem Research School and
by MINECO-FEDER, grants FIS2012-36113-C03-03, FIS2014-55563-REDC, and FIS2015-69295-C3-1-P.
}

\section*{References}
\bibliographystyle{iopart-num}
\providecommand{\newblock}{}

\vfill
\appendix

\section{Geometric Parameters}

The optimized parameters with $f_{g_P}$ \eqref{eq:fgp} and $\widetilde{f}_{g_P}$ \eqref{eq:tfgp}
as objective functions are given in Tab.~\ref{tab:fgp}.
Those with objective functions
$f_{g_\text{far}}$ \eqref{eq:fgfar} and $\widetilde{f}_{g_\text{far}}$ \eqref{eq:tfgfar}
are given in Tab.~\ref{tab:fgfar}.

\begin{table}[h]
	\begin{center}
	\begin{tabular}{c||c|c|c|c||c|c|c|c||c}
		obj.~fct. & $x_1^A$ & $y_1^A$ & $x_2^A$ & $y_2^A$ & $x_1^B$ & $y_1^B$ & $x_2^B$ & $y_2^B$ & $a$
		\\ \hline
		$f_{g_P}$ & 20.00 & 145.51 & 80.00 & 20.00 & 80.00 & 124.85 & 80.00 & 150.51 & 200.00
		\\
		$\widetilde{f}_{g_P}$ & 20.00 & 72.62 & 80.00 & 71.30 & 79.99 & 152.40 & 80.00 & 140.15 & 200.08
	\end{tabular}
	\end{center}
	\caption{Optimized geometric parameters (Fig.~\ref{fig:geo})
	for optimization of relative difference of Purcell factors \eqref{eq:fgp} and \eqref{eq:tfgp}.}
	\label{tab:fgp}
\end{table}

\begin{table}[h]
	\begin{center}
	\begin{tabular}{c||c|c|c|c||c|c|c|c||c}
		obj.~fct. & $x_1^A$ & $y_1^A$ & $x_2^A$ & $y_2^A$ & $x_1^B$ & $y_1^B$ & $x_2^B$ & $y_2^B$ & $a$
		\\ \hline
		$f_{g_\text{far}}$ & 20.00 & 170.00 & 32.51 & 150.31 & 25.62 & 21.57 & 20.19 & 20.00 & 291.43
		\\
		$\widetilde{f}_{g_\text{far}}$ & 35.42 & 186.48 & 62.75 & 123.69 & 56.07 & 176.39 & 77.99 & 159.05 & 196.44 
	\end{tabular}
	\end{center}
	\caption{Optimized geometric parameters (Fig.~\ref{fig:geo})
	for optimization of relative difference of scattered chirality \eqref{eq:fgfar} and \eqref{eq:tfgfar}.}
	\label{tab:fgfar}
\end{table}

\end{document}

%% file: preamble.tex
\usepackage{amsfonts}
\usepackage{amsmath}
\usepackage{color}
\usepackage{bbold}
\usepackage{hhline}
\usepackage[export]{adjustbox}

\newcommand{\epsi}{\varepsilon}
\newcommand{\mue}{\mu}

\newcommand{\Eth}{\vec{\mathcal{E}}}
\newcommand{\Dth}{\vec{\mathcal{D}}}
\newcommand{\Hth}{\vec{\mathcal{H}}}
\newcommand{\Bth}{\vec{\mathcal{B}}}
\newcommand{\Jth}{\vec{\mathcal{J}}}

\newcommand{\Wemi}{W_\text{emi}}

\newcommand{\Wfar}{W_\text{far}}

\newcommand{\Xconv}{X_\text{conv}}

\newcommand{\Xemi}{X_\text{emi}}
\newcommand{\Xfar}{X_\text{far}}

\newcommand{\Ethinc}{\Eth_\text{inc}}
\newcommand{\Hthinc}{\Hth_\text{inc}}
\newcommand{\Bthinc}{\Bth_\text{inc}}

\newcommand{\Ethsca}{\Eth_\text{sca}}
\newcommand{\Hthsca}{\Hth_\text{sca}}
\newcommand{\Bthsca}{\Bth_\text{sca}}

\newcommand{\Sigm}{\vec{\Sigma}}

\newcommand{\chielth}{\mathfrak{X}_\text{e}}
\newcommand{\chimath}{\mathfrak{X}_\text{m}}
\newcommand{\Sigmth}{{\vec{\mathfrak{S}}}}

\renewcommand{\vec}[1]{\boldsymbol{#1}}

\newcommand{\xx}{\vec{x}}

\newcommand{\pp}{\vec{p}}
\newcommand{\mm}{\vec{m}}

\newcommand{\nabl}{\vec{\nabla}}
\newcommand{\rot}[1]{\left( \nabl \times #1 \right)}

\newcommand{\real}[1]{\operatorname{Re}\left(#1\right)}
\newcommand{\imag}[1]{\operatorname{Im}\left(#1\right)}

\usepackage{tikz}
\usetikzlibrary{positioning}
\usetikzlibrary{calc,shapes}